\documentclass[aps,prb,twocolumn,superscriptaddress,showkeys]{revtex4-1}
\usepackage{amsmath,amssymb}
\usepackage{graphicx}
\usepackage{dcolumn}
\usepackage{bm}
\usepackage{esvect}
\usepackage{float}
\usepackage{color}
\usepackage{amsmath}
\usepackage{amsfonts}
\definecolor{rojo}{rgb}{1,0,0}
\definecolor{verde}{rgb}{0,0.8,0.2}
\definecolor{azul}{rgb}{0,0,1}
\definecolor{rosa}{cmyk}{0,1,0,0}
\usepackage[export]{adjustbox}
\usepackage{latexsym}
\usepackage{mathrsfs}
\usepackage{bm}
\usepackage{amsthm}
\usepackage{bbm}
\usepackage{amsfonts}
\usepackage{color}
\usepackage{epsfig}
\usepackage{hyperref}
\hypersetup{colorlinks=true, 
            citecolor=blue,
            urlcolor=black,
            linkcolor=blue}
\usepackage{soul}
\DeclareGraphicsExtensions{.pdf,.png,.jpg,.pdf}
\usepackage{multirow}
\usepackage{changes}
\usepackage{fancyvrb}

\DeclareUnicodeCharacter{0301}{\'{e}}
\newcommand{\nbox}{NbOX$_2$ }

\begin{document}

\title{Photostriction-Driven Phase Transition in Layered Chiral NbOX$_2$ Crystals: \\~Electrical-Field-Controlled Enantiomer Selectivity}

\author{Jorge Cardenas-Gamboa}
\email{jicg1@ifw-dresden.de}
\affiliation{Leibniz Institute for Solid State and Materials Research, IFW Dresden, Helmholtzstraße 20, 01069 Dresden, Germany}

\author{Martin Gutierrez-Amigo}
\affiliation{Department of Applied Physics, Aalto University School of Science, FI-00076 Aalto, Finland}

\author{Aritz Leonardo}
\affiliation{ Donostia International Physics Center, Donostia-San Sebastian 20018 Gipuzkoa, Spain}
\affiliation{EHU Quantum Center, University of the Basque Country UPV/EHU, 48940 Leioa, Spain}

\author{Gregory A. Fiete}
\affiliation{Northeastern University, Boston, Massachusetts 02115, USA}
\affiliation{Quantum Materials and Sensing Institute, Northeastern University, Burlington, Massachusetts 01803, USA}
\affiliation{Department of Physics, Massachusetts Institute of Technology, Cambridge, MA 02139, USA}
\affiliation{Department of Physics, Harvard University, Cambridge, MA 02138, USA}


\author{Juan L. Mañes}
\affiliation{Physics Department, University of the Basque Country (UPV/EHU), Bilbao, Spain}
\affiliation{EHU Quantum Center, University of the Basque Country UPV/EHU, 48940 Leioa, Spain}

\author{Jeroen van den Brink}
\affiliation{Leibniz Institute for Solid State and Materials Research, IFW Dresden, Helmholtzstraße 20, 01069 Dresden, Germany}
\affiliation{Würzburg-Dresden Cluster of Excellence Ct.qmat, Technische Universitat Dresden, 01062, Dresden, Germany}

\author{Claudia Felser}
\affiliation{Max Planck Institute for Chemical Physics of Solids, 01187 Dresden, Germany}

\author{Maia G. Vergniory}
\email{maia.vergniory@usherbrooke.ca}
\affiliation{ Donostia International Physics Center, Donostia-San Sebastian 20018 Gipuzkoa, Spain}
\affiliation{D\'epartement de physique et Institut quantique, Universit\'e de Sherbrooke, Sherbrooke J1K 2R1 QC, Canada}
\affiliation{Regroupement Qu\'eb\'ecois sur les Mat\'eriaux de Pointe (RQMP), Quebec H3T 3J7, Canada}

\date{\today}

\begin{abstract}

Chiral crystals offer an unique platform for controlling structural handedness through external stimuli. However, the ability to select between structural enantiomers remains challenging, both theoretically and experimentally. In this work, we demonstrate a two-step pathway for enantiomer selectivity in layered chiral \nbox (X = Cl, Br, I) crystals based on photostriction-driven phase transitions. {\it Ab-initio} simulations reveal that optical excitation is capable of inducing a structural phase transition in \nbox from the monoclinic ($C2$) ground state to the higher-symmetry ($C2/m$) structure. In the resulting transient high-symmetry state, an applied electric field breaks the residual inversion-symmetry degeneracy, selectively stabilizing one enantiomeric final state configuration over the other. Our results establish a combined optical-electrical control scheme for chiral materials, enabling reversible and non-contact enantiomer selection with potential applications in ultrafast switching, optoelectronics, and chiral information storage.
\end{abstract}

\maketitle

\section{Introduction}

Chirality is a fundamental symmetry concept in physics, chemistry and biology, where an object cannot be geometrically superimposed with its mirror image. In crystalline solids, chirality emerges from structural arrangements that lack inversion or mirror symmetries, producing enantiomers with identical chemical composition but potentially different physical behaviors~\cite{intr_flack2003chiral, intr_fecher2022chirality}. This structural handedness governs diverse phenomena, such as circular dichroism~\cite{intr_jahnigen2023vibrational}, chiral phonon transport~\cite{intro_thingstad2019chiral}, enantioselective catalysis~\cite{intr_wang2025direct}, and chiral-induced spin selectivity~\cite{naaman2012chiral,evers2022theory}, with applications spanning optoelectronics, spintronics, and emerging quantum materials and devices~\cite{dev_crassous2023materials,dev_kumar2020topological,dev_wang2024chiral}. 

One of the major challenges in material science lies in achieving enantioselective control, that is, the ability to reversibly switch between distinct enantiomeric states. Photostriction, the light-induced deformation of crystals, provides a non-contact pathway for inducing symmetry and phase transitions through non-thermal strain generated under illumination~\cite{pht_kundys2015photostrictive}. This optomechanical effect opens new perspectives for integrating light-driven structural control into switchable devices. Photostriction has been extensively investigated in ferroelectrics~\cite{pht_paillard2016photostriction}, multiferroic interfaces~\cite{pht_iurchuk2016optical}, polar semiconductors~\cite{pht_polar_lagowski1972photomechanical,pht_polar_lagowski1974photomechanical}, organic polymers~\cite{pht_orgn_finkelmann2001new}, and materials exhibiting switchable properties~\cite{pht_review_xiang2024high,pht_dansou2024tuning}, just to name a few instances.

The focus of this work is the family of layered niobium oxide dihalides, \nbox (X = Cl, Br, I), which has recently emerged as a versatile platform for exploring the interplay between chirality, ferroelectricity, and external stimuli. These materials crystallize in a monoclinic (C2) structure that is intrinsically chiral, while hosting robust in-plane ferroelectricity persisting from the monolayer limit to the bulk~\cite{nbo_jia2019niobium,nbo_liu2023ferroelectricity}. In addition to their tunable ferro- and antiferroelectric order, \nbox compounds exhibit exceptional optoelectronic characteristics, including strong light absorption in the visible range~\cite{nbo_light_guo2023ultrathin}, high second-harmonic generation efficiencies~\cite{nbo_shg_fu2023manipulating}, and even monolayer-like excitonic behavior in the bulk~\cite{nbo_exc_wang2024indirect}. Their unique combination of tunable ferroic order and pronounced nonlinear optical response makes them attractive not only for polarization-sensitive optoelectronic devices, but also as a direct platform to drive structural phase transitions~\cite{nbo_photostriction_arxiv_dansou2025photoinduced}.

A promising strategy for controlling structural chirality is to exploit materials in which a chiral ground state lies energetically close to an achiral configuration~\cite{intr_fecher2022chirality}. Gutierrez-Amigo \textit{et al.}~\cite{nbo_martin_arxiv_gutierrez2025emergent} recently demonstrated this mechanism in the \nbox family, where two enantiomorphic $C2$ ground states are linked through a low-energy achiral $C2/m$ phase. The relatively small barrier between the chiral ground state and this intermediate achiral configuration provides a natural pathway for such symmetry control. This intermediate state can be stabilized by anharmonic effects or external pressure, while the ultimate handedness can be selectively imposed by a symmetry-breaking electric field. 

In this work, we introduce a photostriction-driven mechanism to control structural chirality in bulk \nbox crystals. We propose a two-step process: (1) photoexcitation to generate a structural deformation that drives the system into the achiral $C2/m$ phase, and (2) an external electric field to steer the relaxation pathway towards a chosen enantiomer, thereby establishing a fully optoelectrical protocol for enantiomer selection in layered chiral ferroelectrics, as illustrated in Fig.~\ref{figure1}(a). To validate this mechanism, we compute the \textit{shift current} response, demonstrating: (i) the mirror-symmetric relation between the two enantiomeric ground states and (ii) how this nonlinear effect can be exploited as a chiral-selective control mechanism.

Our article is organized as follows. First, we describe the {\em ab initio} computational methods and structural properties for \nbox crystals. Next, we examine the photostriction-induced phase transition, supported by phonon and band structure calculations to evaluate phase stability. We then analyze the effect of an external electric field in selecting a particular enantiomeric final state and assess the corresponding shift-current response. Finally, we summarize the main findings and discuss their implications for controlling chirality in ferroelectric materials.

\section{Computation details} \label{computational}


All calculations were performed within density functional theory (DFT) using the Quantum ESPRESSO package (version 7.3)~\cite{DFT_QE_giannozzi2009quantum,DFT_QE_giannozzi2017advanced} with Optimized Norm-Conserving Vanderbilt pseudopotentials including valence electrons. A plane-wave cutoff of 50 Ry was used for the wavefunctions and 400 Ry for the charge density, together with a Methfessel–Paxton~\cite{DFT_tethfese:methfessel1989high} smearing of 0.02 Ry. The Brillouin-zone integrations employed an $8 \times 8 \times 5$ Monkhorst–Pack $k$-point mesh. Structural relaxations were carried out until atomic forces were smaller than $10^{-4}$ Ry/Bohr and stress components were below 0.1 kbar. Exchange–correlation effects were treated within the generalized gradient approximation (GGA) using the Perdew–Burke–Ernzerhof (PBE) functional~\cite{DFT_PBE_perdew1996generalized}. Interlayer interactions were accounted for through van der Waals corrections using the semi-empirical DFT-D2 scheme of Grimme \textit{et al.}~\cite{DFT_Grimm_grimme2006semiempirical}, which ensures the correct energetic ordering of phases and improves agreement between calculated and experimental lattice parameters (see Supplementary I).

To model laser irradiation, we adopted the scheme proposed in Ref.~\cite{QE_marini2021lattice,QE_mocatti2023light} for semiconductors and insulators, in which photocarrier excitation is simulated by promoting a fraction $x$ of valence electrons to conduction-band states, corresponding to a photocarrier concentration $n_e = x$ $e^{-}$/u.c., where u.c. denotes the unit cell. Harmonic phonon spectra were obtained within density functional perturbation theory (DFPT) on a $2 \times 2 \times 2$ $q$-point mesh for both the $C2/m$ and $C2$ phases. Spin–orbit coupling (SOC) was neglected in these calculations, and phonon dispersions were generated by Fourier interpolation. Ferroelectric properties, were computed using the modern theory of polarization (Berry-phase formalism) as implemented in Quantum Espresso~\cite{DFT_QE_berry_scandolo2005first}.

Shift-current responses were computed using the Wannierization approach implemented in Wannier90~\cite{Wannier90_mostofi2014updated} and WannierBerri~\cite{Wannierberri_tsirkin2021high,Wannierberri_shift_current_ibanez2018ab}. The tight-binding model was constructed by projecting onto O-$p$, X-$p$ (X = Cl, Br, I), and Nb-$d$ orbitals, centered on their respective atomic sites.


\section{Structural and electronic properties of \nbox} \label{section2}

\begin{figure*}[t]
    \centering
    \includegraphics[scale=0.90]{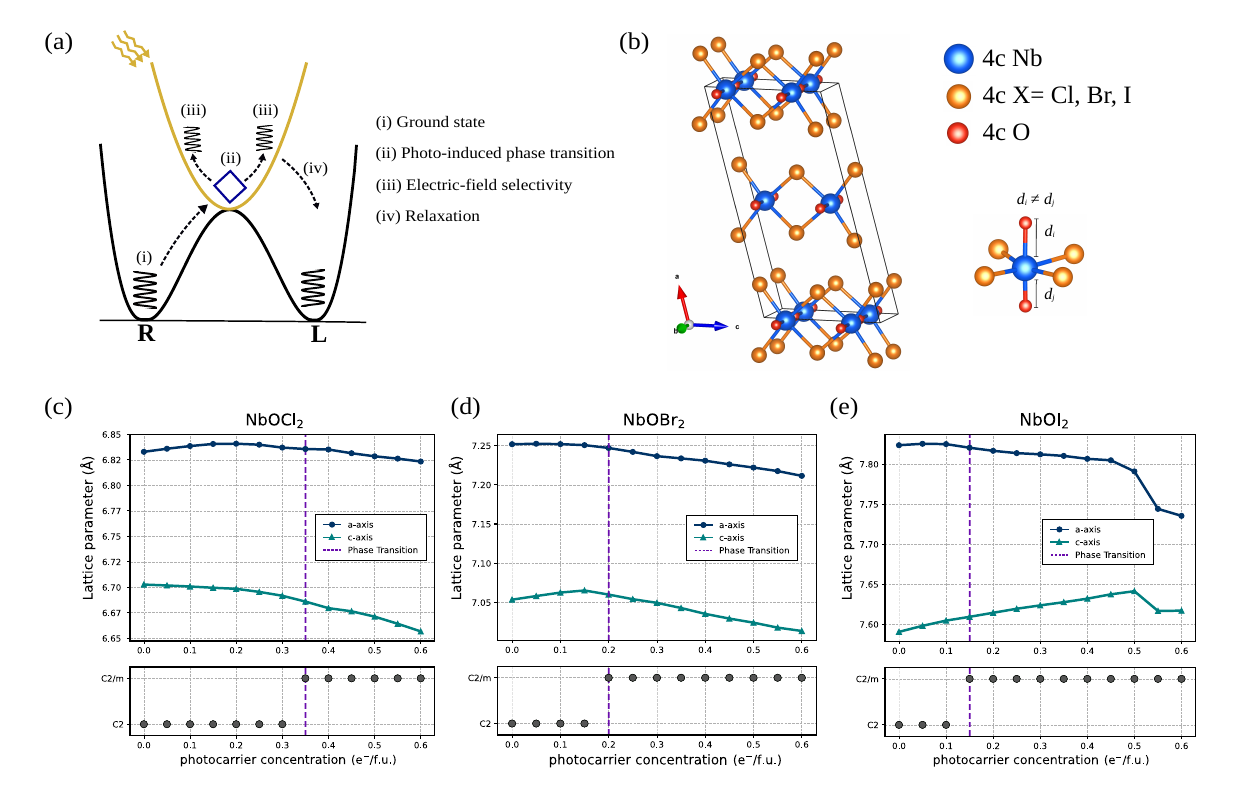}
    \caption{
    (a) Schematic view of the two-step crystal chirality selection mechanism proposed: (1) light-induced nonthermal phase transition and (2) electric-field–driven enantiomer stabilization. (b) Layered structure of NbOX$_2$ (X = Cl, Br, I), with Nb (blue), O (red), and X (yellow) atoms; $d_i$ and $d_j$ denote Nb–O bond lengths. (c–e) Evolution of lattice constants (upper panels) and space group (lower panels) as a function of the photocarrier concentration for (c): NbOCl$_2$, (d): NbOBr$_2$, and (e): NbOI$_2$, respectively. Purple line represent the quantity required for a symmetry-phase transition.
    }
    \label{figure1}
\end{figure*}

The family of van der Waals (vdW) materials \nbox (X = Cl, Br, I) consists of stacked \nbox monolayers along the out-of-plane $a$-axis. As illustrated in Fig.\ref{figure1}(b), within each monolayer the Nb atoms are arranged in chains of edge-sharing distorted octahedra, interconnected through O atoms along the $b$-direction. A displacement of atoms along the $c$-axis leads to a Peierls distortion, accompanied by a metal–to-insulator transition\cite{nbox_theoretical_ye2023manipulation}. In addition, a second-order Peierls distortion along the $b$-axis displaces Nb atoms off-center within the \nbox octahedra toward the corner O atoms. This distortion produces two inequivalent Nb–O bond lengths and induces in-plane spontaneous polarization (ferroelectricity) along the $b$-direction. These structural features are depicted in Fig.~\ref{figure1}(b) (octahedra enviroment), where the two inequivalent Nb–O bonds $(d_i, d_j)$ are highlighted. Together, these distortions stabilize a monoclinic structure with space group $C2$ along the b-axis.

To further characterize \nbox we performed DFT calculations of its band structure and density of states (DOS), following the description in Ref.~\cite{nbox_theoretical_ye2023manipulation,nbo_liu2023ferroelectricity} (see Supplementary III). All structures exhibit a finite band gap, with contributions primarily from Nb-4$d$ and X-$p$ orbitals. A nearly flat band appears close to the Fermi level $E_F$, arising from the localized $d_{z^2}$ orbitals of the underlayer Nb atoms. Within a crystal field picture, the Nb-$d$ manifold splits into three non-degenerate orbitals ($d_{z^2}$, $d_{xz}$, $d_{x^2-y^2}$) and a doubly degenerate pair ($d_{xy}$, $d_{yz}$). Given the valence configuration of Nb ($4d^45s^1$), each Nb$^{4+}$ cation contributes a single unpaired 4$d$ electron, which half-fills the $d_{z^2}$ orbital. Dimerization of Nb atoms along the $c$-axis, corresponding to a Peierls distortion, further splits the $d_{z^2}$ states into bonding- and antibonding-like levels. The bonding branch forms an isolated flat band near the top of the valence band (VB), while the antibonding counterpart is shifted to about 1 eV above the conduction band (CB) minimum. In addition, O-2$p$ and X-$p$ orbitals contribute mainly to the VB at lower energies, remaining well separated from the flat Nb-$d_{z^2}$ band.

\begin{figure*}[t]
    \centering
    \includegraphics[scale=0.90]{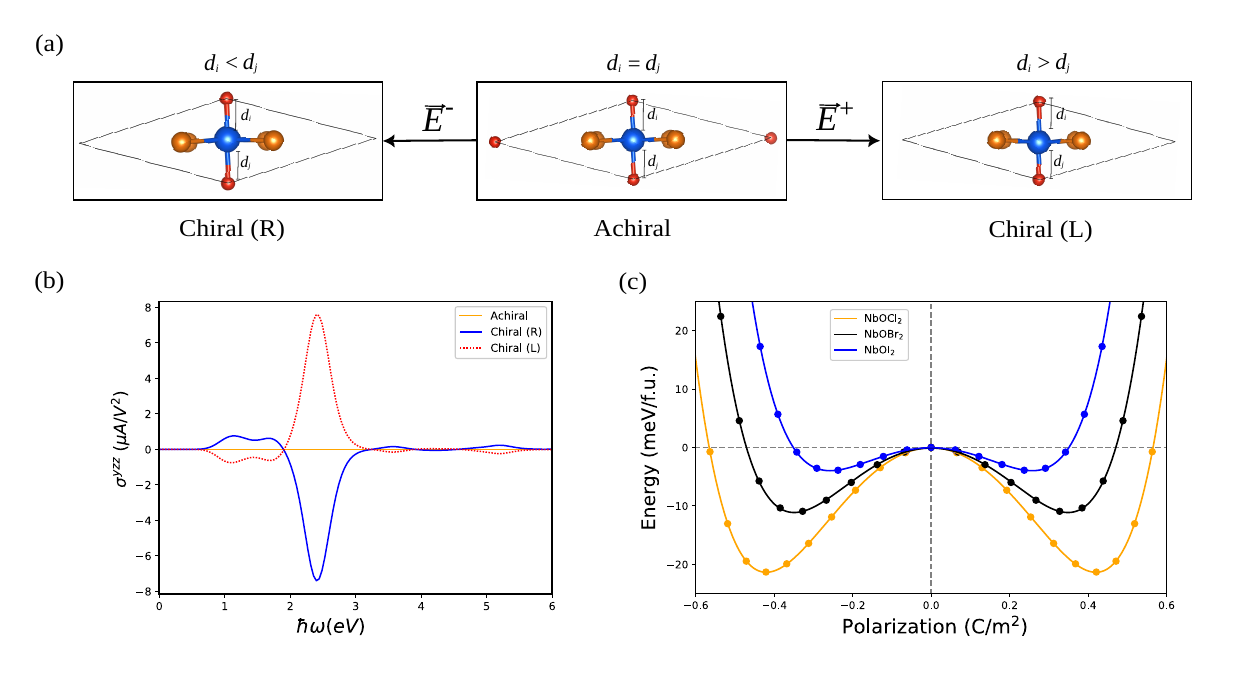}
    \caption{
    (a) Schematic illustration (along the c-axis) of the electric-field–selective mechanism, represented in terms of right- and left-handed structures defined by the Nb–O bond lengths ($d_i$, $d_j$). (b) Calculated shift-current spectra as a function of photon energy for the structures in (a): achiral (orange), right-handed (dashed red), and left-handed (solid blue). (c) Energy–polarization double-well profiles for ferroelectric bulk NbO$X_2$. Dots denote the calculated data, while solid lines represent fits based on the Landau model (see Supplementary V).
    }
    \label{figure3_chiralityinduced}
\end{figure*}

\section{Photo-induced phase transitions}

Having established the chiral ground state, we now turn to the effect of photoexcitation in these materials. Upon above-bandgap excitation, absorbed photons generate a non-equilibrium carrier distribution that instantaneously modifies the electronic density and, consequently, the Born–Oppenheimer potential energy surface experienced by the nuclei. In \nbox this redistribution couples strongly to polar lattice degrees of freedom through both the piezoelectric response and the electronic (photostrictive) pressure. The resulting stress is highly anisotropic, leading to differential changes in key Nb–O bond lengths ($d_i$ and $d_j$) within the layers, as well as in the lattice parameters, thereby reshaping the Born–Oppenheimer potential landscape, as illustrated in Fig.~\ref{figure1}(a) (yellow line).

We computed the lattice parameters and space group evolution of \nbox as a function of photocarrier concentration, as shown in Fig.\ref{figure1}(c–d). All symmetry phases (bottom) were identified using the Spglib package\cite{spglib_togo2024spglib} with a tolerance of 0.002~\AA. At low excitation densities, the induced strain remains within the elastic regime, preserving the original chiral $C2$ symmetry. However, beyond a critical carrier density $n_{e}$ (dashed purple line), photostrictive deformation drives a cooperative lattice instability, selectively softening phonon modes associated with the Nb–O bonding network. This leads to a higher-symmetry achiral phase ($C2/m$), consistent with the mechanism proposed in Ref.\cite{nbo_martin_arxiv_gutierrez2025emergent}. Phonon and band-structure calculations were performed for both the ground state and the photo-induced phases under dark conditions, confirming the stability of the chiral phase and showing that the higher-symmetry structure exhibits unstable phonon modes and is therefore dynamically unstable (see Supplementary II).

To estimate the laser fluence required to excite the critical photocarrier concentrations ($n_e \approx 0.15$–$0.35$ $e^-/\text{f.u.}$), we estimate the required laser fluence $I_0$ needed to generate this carrier concentration by relating it to the material's fundamental optical properties, as~\cite{laser_for_paillard2023light,laser_for_peng2022tunable}:

\begin{equation}
n_e = \frac{I_0 (1 - R) \alpha \Omega_0}{\hbar\omega},
\end{equation}
where $R$ is the reflectivity, $\alpha$ is the absorption coefficient, $\Omega_0$ is the unit cell volume, and $\hbar\omega$ is the photon energy. We computed $\alpha$ and $R$ from the frequency-dependent dielectric tensor (see Supplementary Material IV).

The results, summarized in Table~\ref{laser}, show that the fluences required to reach the critical carrier density $n_e$ span from $1.55~\text{mJ}/\text{cm}^2$ to $25.30~\text{mJ}/\text{cm}^2$ for bulk \nbox. These values fall well within the operational range of standard ultrafast laser systems, confirming the practical feasibility of the photoexcitation step in our proposed mechanism~\cite{review_chen2021photostrictive}. Notably, the strong increase in the absorption coefficient $\alpha$ along the Cl–Br–I series leads to a pronounced reduction in the required fluence, highlighting NbOI$_2$ as the most favorable candidate for experimental realization.

\begin{table}[h]
\caption{Estimated laser fluence $(I_0)$ required to drive the photoinduced phase transition. The critical carrier density $n_e$ is determined from the structural calculations in Fig.~\ref{figure1}(c-d). The absorption coefficient $\alpha$ and reflectivity $R$ are calculated averages at 2.0 eV. The unit cell volume $\Omega_0$ is computed from the $C2$ ground state.}
\label{laser} 
\centering
\begin{tabular}{lccccc}
\hline
Material & \multicolumn{1}{c}{$n_e$} & \multicolumn{1}{c}{$\alpha$} & \multicolumn{1}{c}{$\Omega_0$} & \multicolumn{1}{c}{$R$} & \multicolumn{1}{c}{$I_0$} \\ 
         & \multicolumn{1}{c}{(e$^{-}$/f.u.)} & \multicolumn{1}{c}{(cm$^{-1}$)} & \multicolumn{1}{c}{(\AA$^{3}$)} & \multicolumn{1}{c}{} & \multicolumn{1}{c}{(mJ/cm$^{2}$)} \\
\hline
NbOCl$_2$ & 0.35 & $4.11 \times 10^{4}$ & 164.63 & 0.345 & 25.30 \\ 
NbOBr$_2$ & 0.20 & $6.45 \times 10^{4}$ & 184.17 & 0.387 & 8.78 \\ 
NbOI$_2$  & 0.15 & $2.612 \times 10^{5}$ & 215.41 & 0.450 & 1.55 \\ 
\hline
\end{tabular}
\end{table}

\section{Electric Field Effects on Enantiomer Selectivity}

The use of external electric fields to control symmetry has been widely explored, particularly in field-induced alignment and orientation of atomic and molecular systems. In these studies, the reflection symmetry was broken by converting an initially aligned population of atoms or molecules into a transverse orientation, establishing one of the earliest paradigms for field-driven symmetry breaking~\cite{electric1_deepa2014chiral,electric2_del2006role}. We extend this principle to induce a preferred handedness by selectively stabilizing one enantiomer depending on the sign of the applied electric field $\mathbf{E}$. 

As discussed in the previous section, reaching the critical photocarrier concentration drives the system into a centrosymmetric (achiral) structure characterized by uniform Nb–O bond lengths (see Fig.~\ref{figure3_chiralityinduced}(a)). Thermodynamically, once the laser is turned off, this achiral state can relax into either of the two enantiomers with equal probability. Previous work~\cite{nbo_martin_arxiv_gutierrez2025emergent} suggests that the relaxation pathway proceeds via the $\Gamma_1^{-}$ mode into the chiral phase; however, an external bias (i.e., an applied electric field) can favor one enantiomer over the other. To test this idea, we simulated the NbOBr$_2$ structure by quenching the photostrictive deformation (turning off the laser) and relaxing the system under an external electric field $\mathbf{E}$ applied along the $b^*$ reciprocal direction with an intensity of $10^{-7}$~a.u. ($0.514$~kV/cm). As shown in Fig.~\ref{figure3_chiralityinduced}(a), the direction of the electric field ($\mathbf{E}$) breaks mirror symmetry and biases the relaxation: a positive field drives the Nb atoms toward bond $d_j$ ($\mathbf{E}^{+}$), stabilizing the left-handed enantiomer, while a negative field favors bond $d_i$ ($\mathbf{E}^{-}$), stabilizing the right-handed enantiomer. This switching simultaneously reverses the polarization direction, in agreement with Born–Oppenheimer simulations under applied electric fields reported in Ref.~\cite{nbo_martin_arxiv_gutierrez2025emergent}.

To probe the nature of the relaxed structures (achiral, right-handed, and left-handed), we calculated the shift current density ($J^{a}$), a second-order nonlinear optical response, under an external electric field $\mathbf{E}$ of frequency $\omega$, following Refs.~\cite{Wannierberri_shift_current_ibanez2018ab,Wannierberri_tsirkin2021high}:

\begin{equation}
j^{a} = 2\sigma^{abc}(0;\omega,-\omega),\mathrm{Re}\left[\mathcal{E}{b}(\omega)\mathcal{E}{c}(-\omega)\right],
\label{shift-current}
\end{equation}
where $\sigma^{abc}(0;\omega,-\omega)$ is the third-rank shift-current response tensor, which transforms analogously to the piezoelectric tensor. 

By symmetry, the shift current vanishes in centrosymmetric systems. For the non-centrosymmetric $C2$ space group, group-theoretical analysis using the MTENSOR program (Bilbao Crystallographic Server~\cite{bilbao1_aroyo2011crystallography,bilbao2_gallego2019automatic}) identifies eight independent nonzero tensor components. To highlight the enantioselective response, we focus on the $\sigma^{yzz}$ component (evaluated at the Fermi level), shown in Fig.~\ref{figure3_chiralityinduced}(b) for the achiral, left-handed, and right-handed cases. As expected, the achiral configuration shows no shift-current response, whereas the two enantiomers display responses of equal magnitude but opposite sign. This opposite behavior, obtained from the relaxed structures after field-induced symmetry breaking, provides direct evidence that the two states are genuine mirror-related enantiomers and confirms the robustness of our combined photoexcitation–electric field approach for engineering and detecting chirality.

It is well established that ferroelectrics enable polarization reversal under an applied electric field, meaning that, in principle, a sufficiently strong field can switch both the polarization and, consequently, the chirality of these systems. To quantify this effect in bulk \nbox we performed Berry-phase polarization calculations and extracted the coercive field $E_c$ (see Supplementary V) from the energy–polarization double-well profiles shown in Fig.~\ref{figure3_chiralityinduced}(c). The resulting switching barrier corresponds to a coercive field of $\sim$100 kV/cm, consistent with values reported experimentally for layered ferroelectrics~\cite{nbo_jia2019niobium}. While such fields are experimentally accessible, they are relatively large, posing challenges for reliable device integration and long-term operation. By contrast, our combined photoexcitation–electric field scheme reduces the required switching field by more than an order of magnitude. Specifically, we demonstrate polarization reversal with fields as small as $\sim0.52$~kV/cm, representing nearly two orders of magnitude reduction relative to the intrinsic bulk coercive field. This dramatic decrease highlights the efficiency of our approach: the optical excitation softens the energy landscape, and the subsequent electric field biases the system toward one enantiomer, enabling deterministic chirality selection. Beyond lowering energy costs, this mechanism opens the door to fast, reversible, and non-invasive control of ferroelectric chirality, making it highly attractive for next-generation applications in ultrafast memory, chiral electronics, and optoelectronic devices.

\section{Conclusions}

In this work, we have proposed and demonstrated a combined optical–electrical strategy to selectively stabilize a preferred handedness in \nbox layered materials. By photoexciting the system, we identified critical photocarrier concentrations that drive a chiral-to-achiral transition. Building on this, we showed that an applied electric field can break mirror symmetry in a controllable manner, with the field polarity determining the favored enantiomer. The computed shift-current response confirmed that the resulting left- and right-handed structures are true mirror-related enantiomers. Finally, Berry-phase calculations revealed that our approach significantly reduces the coercive field required for polarization reversal compared to bulk switching, highlighting a clear experimental advantage. Overall, these results establish photostriction-assisted electric-field control as a powerful platform for tuning structural chirality, opening opportunities for exploiting chiral degrees of freedom in ferroelectric materials with potential applications in devices. 

\section{Acknowledgements} 
M.G.V. thanks support to PID2022-142008NB-I00 funded by  MICIU/AEI/10.13039/501100011033 and FEDER, UE, the Canada Excellence Research Chairs Program for Topological Quantum Matter and to Diputaci\'on Foral de Gipuzkoa Programa Mujeres y Ciencia. This work has been funded by the IKUR Strategy under the collaboration agreement between Ikerbasque Foundation and DIPC on behalf of the Department of Education of the Basque Government and the Deutsche Forschungsgemeinschaft (DFG, German Research Foundation), through the Würzburg-Dresden Cluster of Excellence on Complexity and Topology in Quantum Matter, ct.qmat (EXC 2147, Project ID 390858490) and the GA 3314/1-1 – FOR 5249 (QUAST). A.E. acknowledges funding from the Fonds zur Förderung der Wissenschaftlichen Forschung (FWF) under Grant No. I 5384.  G.A.F. acknowledges funding from the National Science Foundation through DMR-2114825 and additional support from the Alexander support from the Alexander
von Humboldt Foundation.
The work of JLM has been partly supported by the Basque Government Grant No. IT1628-22 and the PID2021-123703NB-C21 grant funded by MCIN/AEI/10.13039/501100011033/ and ERDF; ``A way of making Europe”. 
We acknowledge Iñigo Robredo, Aitzol García-Etxarri and Diego García Ovalle for discussions.

\clearpage

\bibliography{acs-achemso}

\end{document}


\title{Supplementary Material for: \\ Photostriction-Driven Phase Transition in Layered Chiral NbOX$_2$ Crystals: \\~Electric-Field-Controlled Enantiomer Selectivity}

\author{Jorge Cardenas-Gamboa}
\email{jicg1@ifw-dresden.de}
\affiliation{Leibniz Institute for Solid State and Materials Research, IFW Dresden, Helmholtzstraße 20, 01069 Dresden, Germany}

\author{Martin Gutierrez-Amigo}
\affiliation{Department of Applied Physics, Aalto University School of Science, FI-00076 Aalto, Finland}

\author{Aritz Leonardo}
\affiliation{ Donostia International Physics Center, Donostia-San Sebastian 20018 Gipuzkoa, Spain}
\affiliation{EHU Quantum Center, University of the Basque Country UPV/EHU, 48940 Leioa, Spain}

\author{Gregory A. Fiete}
\affiliation{Northeastern University, Boston, Massachusetts 02115, USA}
\affiliation{Quantum Materials and Sensing Institute, Northeastern University, Burlington, Massachusetts 01803, USA}
\affiliation{Department of Physics, Massachusetts Institute of Technology, Cambridge, MA 02139, USA}
\affiliation{Department of Physics, Harvard University, Cambridge, MA 02138, USA}


\author{Juan L. Mañes}
\affiliation{Physics Department, University of the Basque Country (UPV/EHU), Bilbao, Spain}
\affiliation{EHU Quantum Center, University of the Basque Country UPV/EHU, 48940 Leioa, Spain}

\author{Jeroen van den Brink}
\affiliation{Leibniz Institute for Solid State and Materials Research, IFW Dresden, Helmholtzstraße 20, 01069 Dresden, Germany}
\affiliation{Würzburg-Dresden Cluster of Excellence Ct.qmat, Technische Universitat Dresden, 01062, Dresden, Germany}

\author{Claudia Felser}
\affiliation{Max Planck Institute for Chemical Physics of Solids, 01187 Dresden, Germany}

\author{Maia G. Vergniory}
\email{maia.vergniory@usherbrooke.ca}
\affiliation{ Donostia International Physics Center, Donostia-San Sebastian 20018 Gipuzkoa, Spain}
\affiliation{D\'epartement de physique et Institut quantique, Universit\'e de Sherbrooke, Sherbrooke J1K 2R1 QC, Canada}
\affiliation{Regroupement Qu\'eb\'ecois sur les Mat\'eriaux de Pointe (RQMP), Quebec H3T 3J7, Canada}

\date{\today}

\maketitle

\clearpage

\section{Comparison of van der Waals treatments}
\begin{figure}[h]
    \centering
    \includegraphics[scale=0.80]{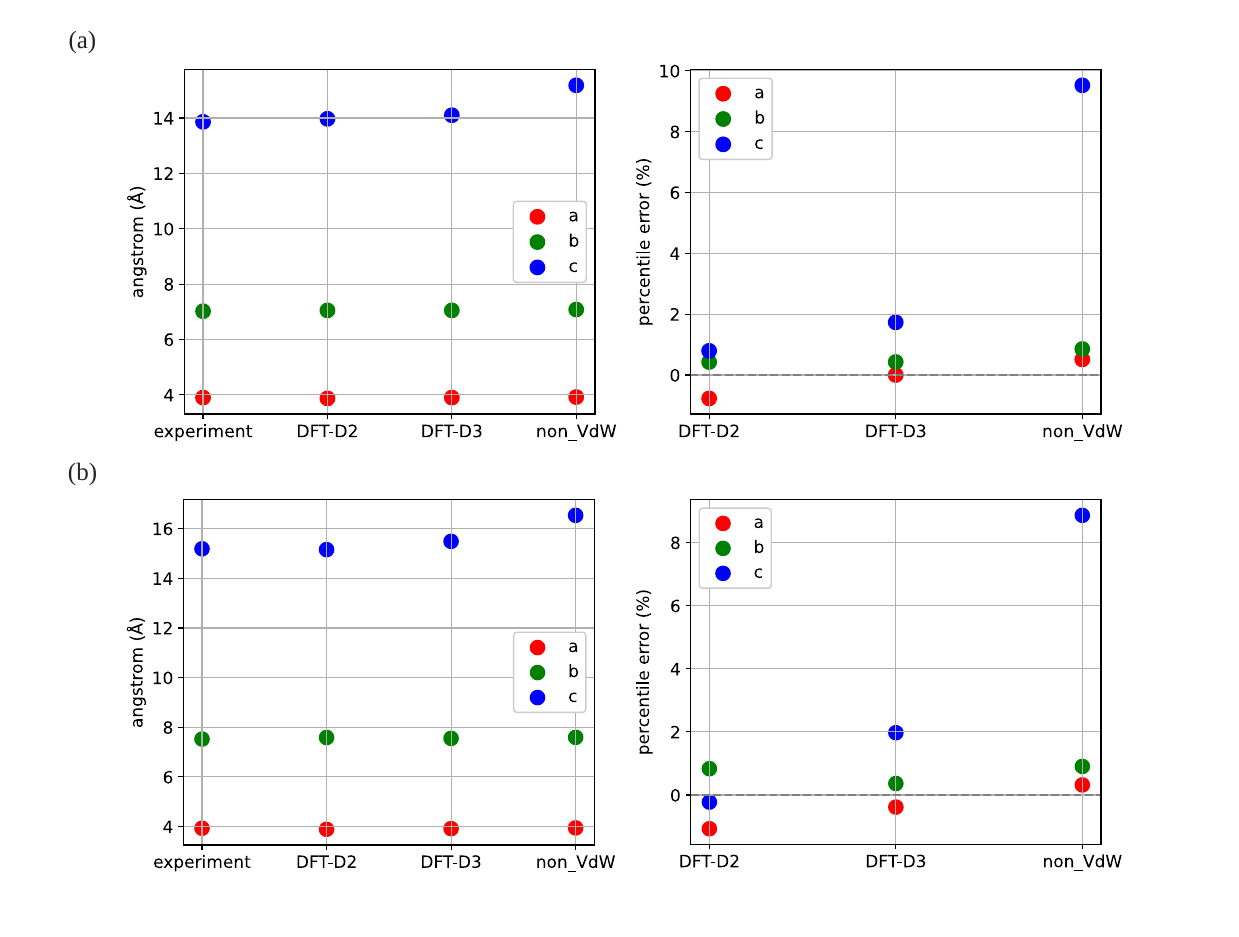}
    \caption{Comparison of van der Waals (vdW) treatments for (a, left) NbOBr$_2$ and (b, left) NbOI$_2$. Experimental lattice parameters are taken from Ref.~\cite{dataexp_shg_fu2023manipulating}. Panels (a, right) and (b, right) show the relative percentage error of each vdW scheme with respect to experiment.}
    \label{experimental_test}
\end{figure}

Given the layered nature of \nbox (Fig.~1(a)), we tested several approaches for incorporating van der Waals (vdW) interactions in structural relaxations. Calculations were performed without spin–orbit coupling (SOC), starting from the experimental lattice parameters reported in Ref.~\cite{dataexp_shg_fu2023manipulating}. We benchmarked three schemes: (i) Grimme’s empirical DFT-D2 correction~\cite{D2_grimme2006semiempirical}, (ii) Grimme’s DFT-D3 correction~\cite{D3_grimme2011effect}, and (iii) a reference case without vdW corrections. The resulting relaxed lattice parameters were compared against experimental data (see Fig.~S\ref{experimental_test}). Among the tested schemes, DFT-D2 gave the best agreement, particularly along the interlayer axis, confirming its suitability for describing vdW interactions in this family. Therefore, all subsequent calculations employed the DFT-D2 correction.

\clearpage

\section{Band Structures and Phonon Spectra at Different Photocarrier Concentrations} \label{computational}

\begin{figure}[h]
    \centering
    \includegraphics[scale=0.80]{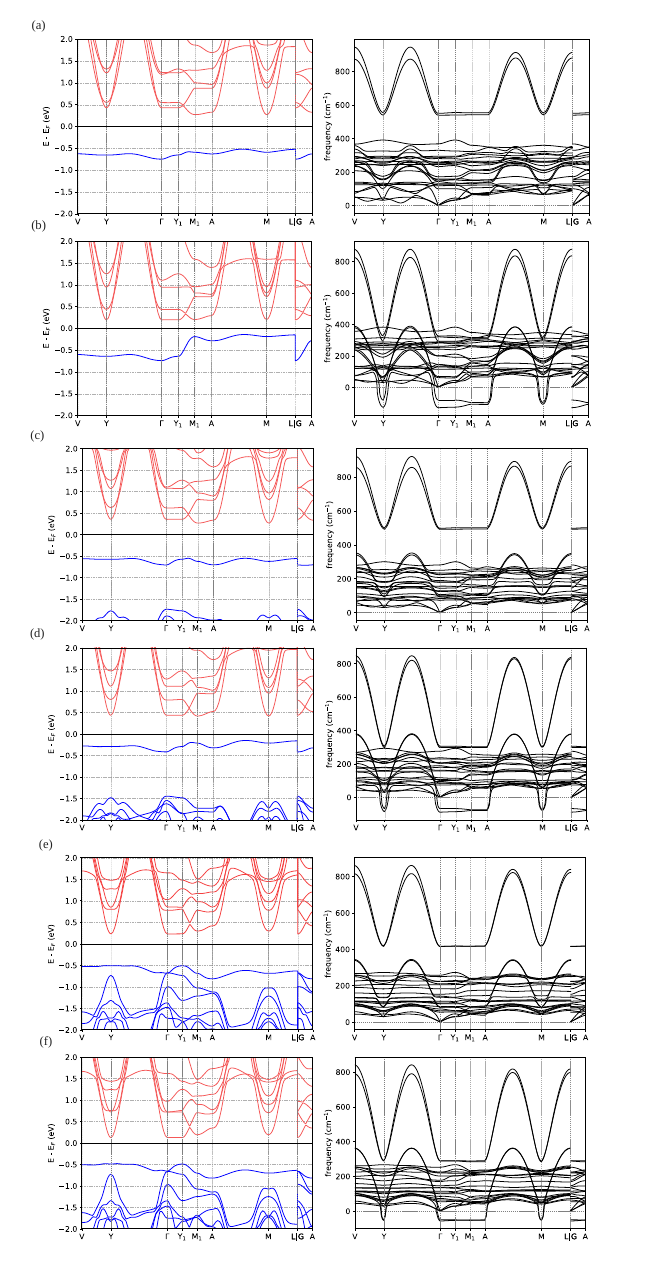}
    \caption{Electronic band structures and phonon spectra of \nbox\ at two representative photocarrier concentrations. 
    (a,b) NbOCl$_2$ at $n_e = 0$ (ground state) and $n_e = 0.35$~e$^-$/f.u., respectively. 
    (c,d) NbOBr$_2$ at $n_e = 0$ and $n_e = 0.20$~e$^-$/f.u. 
    (e,f) NbOI$_2$ at $n_e = 0$ and $n_e = 0.15$~e$^-$/f.u. 
    In all cases, $n_e = 0$ corresponds to the equilibrium monoclinic phase, while finite carrier concentrations illustrate the critical regimes driving the chiral-to-achiral transition.}
    \label{band_and_phonon}
\end{figure}

For the structures analyzed in Fig.~1(c–e), we report the corresponding electronic band structures and phonon spectra (calculated under dark conditions) at two representative photocarrier concentrations: (i) $n_e = 0$, corresponding to the ground state, and (ii) $n_e = n_c$, where $n_c$ denotes the critical carrier density required to drive the chiral-to-achiral ($C2/m$) phase transition (see dashed purple line in Fig.~2(c–e)). The results, presented in Fig.~S\ref{band_and_phonon}, show that in the achiral phase all \nbox compounds exhibit phonon instabilities, confirming that this structure is dynamically unstable.

\section{Projected Band-Structure}

\begin{figure}[H]
    \centering
    \includegraphics{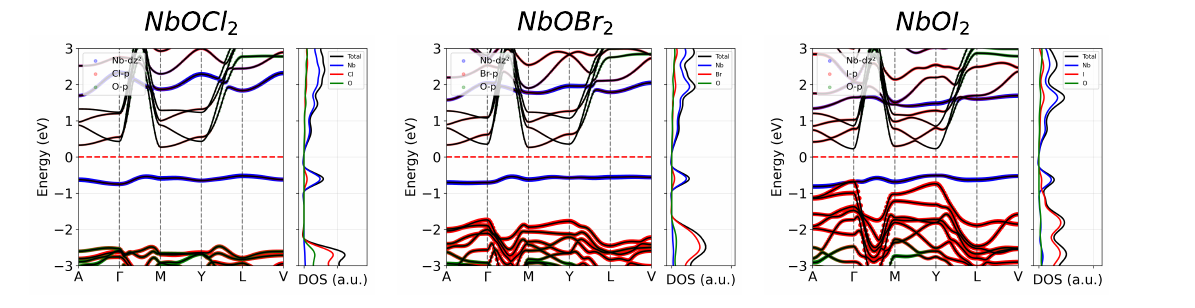}
    \caption{Orbital-resolved band structure and density of states (DOS) of bulk \nbox in the monoclinic ($C2$) ground state. The contribution of the Nb-$d_{z^{2}}$ orbital is highlighted by blue points in the band structure.}

    \label{optical_properties}
\end{figure}

\section{Optical absorption of bulk \nbox}
%
%
\begin{figure}[H]
    \centering
    \includegraphics[scale=0.85]{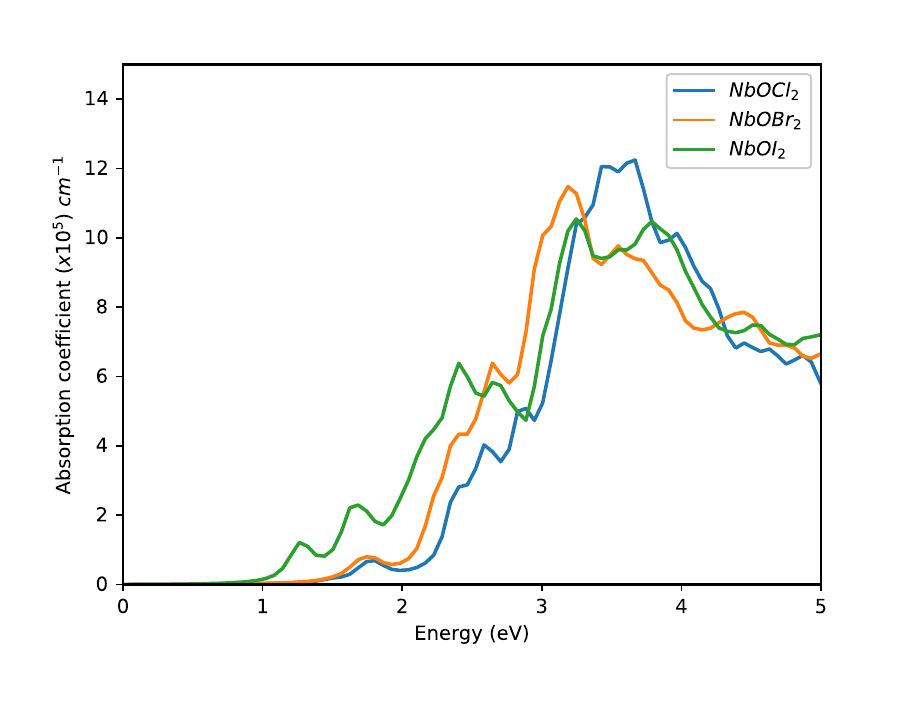}
    \caption{Calculated optical absorption coefficient $(\alpha)$ of ferroelectric \nbox\ bulk, obtained using the DFT-D2 functional. 
    The spectra highlight the energy ranges relevant for photoexcitation processes and allow comparison of absorption edges across different halides. 
    These results provide the basis for estimating the photocarrier concentrations required to drive the chiral-to-achiral transition.}

    \label{optical_properties}
\end{figure}

\section{Landau Model for Ferroelectric \nbox}

The ferroelectric behavior of \nbox\ can be described within the Landau free-energy expansion~\cite{landau1_lifshitz1980landau,landau2_hohenberg2015introduction}:

\begin{equation}
    F(P) = \tfrac{1}{2}aP^{2} + \tfrac{1}{4}bP^{4} + \tfrac{1}{6}cP^{6} - EP,
    \label{eq1}
\end{equation}
where $P$ is the polarization, $a$, $b$, and $c$ are expansion coefficients, and $E$ is the applied electric field.  
The coefficients were obtained by fitting the Berry-phase polarization energies to Eq.~\ref{eq1}, as shown in Fig.~2(c).

The equilibrium polarization corresponds to the minima of $F(P)$, determined by the stationary condition

\begin{equation}
    \frac{\partial F}{\partial P} = 0,
    \label{eq2}
\end{equation}
which leads to the constitutive relation

\begin{equation}
    E = aP + bP^{3} + cP^{5}.
    \label{eq3}
\end{equation}

\begin{figure}[H]
    \centering
    \includegraphics[scale=0.85]{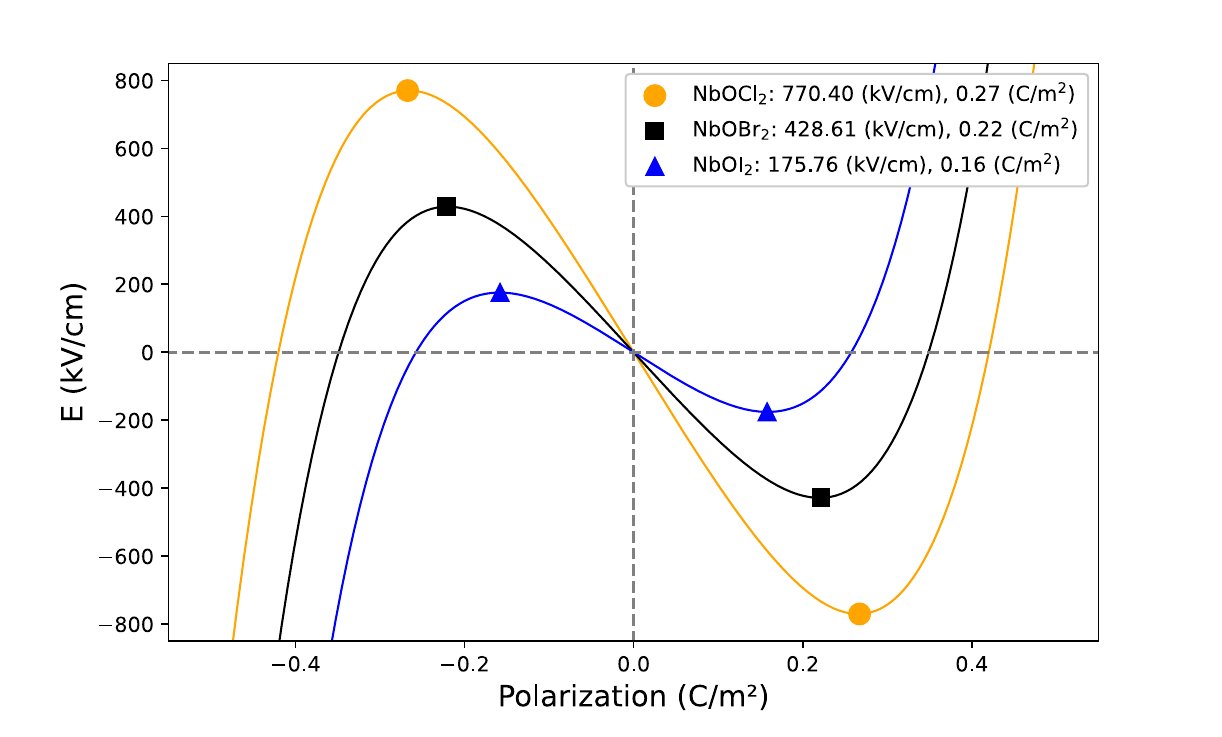}
    \caption{Polarization–electric field ($P$–$E$) characteristics of bulk \nbox\ derived from the Landau free-energy expansion. The coercive field $E_c$ and spontaneous polarization values extracted from the fits are indicated in the legend.}
    \label{coercive}
\end{figure}

Solving Eq.~\ref{eq3} yields the polarization–electric field ($P$–$E$) characteristics. From these, we extracted the coercive field $E_c$ for bulk \nbox, i.e., the minimum external field required to reverse the spontaneous polarization. The computed $P$–$E$ loops and the corresponding coercive fields are presented in Fig.~S\ref{coercive}.  

\bibliography{acs-achemso_supp}